\documentclass[preprint,12pt]{elsarticle}
\usepackage{lineno}
\modulolinenumbers[1]
\usepackage{geometry}
\usepackage{graphicx}
\usepackage[usenames,dvipsnames]{color}
\usepackage[english]{babel}
\usepackage{epstopdf}
\usepackage{latexsym}
\usepackage{url}
\usepackage{amstext}
\usepackage{amssymb}
\usepackage{amsmath,amsthm}
\usepackage{pifont}
\usepackage{ulem}
\usepackage[pagebackref=false]{hyperref}
\hypersetup{
    colorlinks=true,
    linkcolor=blue,
    filecolor=magenta,
    urlcolor=blue,
}

\biboptions{sort&compress}

\begin{document}

\begin{frontmatter}

\title{Adaptive Modelling of Anti-tau Treatments for Neurodegenerative Disorders Based on the Bayesian Approach with Physics-Informed Neural Networks}

\author[inst1]{Swadesh Pal}
\ead{spal@wlu.ca}
\author[inst1,inst2]{Roderick Melnik}
\ead{rmelnik@wlu.ca}

\address[inst1]{MS2 Discovery Interdisciplinary Research Institute, Wilfrid Laurier University, Waterloo, Canada}
\address[inst2]{BCAM - Basque Center for Applied Mathematics, Bilbao, Spain}

\begin{abstract}
Alzheimer's disease (AD) is a complex neurodegenerative disorder characterized by the accumulation of amyloid-beta (A$\beta$) and phosphorylated tau (p-tau) proteins, leading to cognitive decline measured by the Alzheimer's Disease Assessment Scale (ADAS) score. In this study, we develop and analyze a system of ordinary differential equation models to describe the interactions between A$\beta$, p-tau, and ADAS score, providing a mechanistic understanding of disease progression. To ensure accurate model calibration, we employ Bayesian inference and Physics-Informed Neural Networks (PINNs) for parameter estimation based on Alzheimer's Disease Neuroimaging Initiative data. The data-driven Bayesian approach enables uncertainty quantification, improving confidence in model predictions, while the PINN framework leverages neural networks to capture complex dynamics directly from data. Furthermore, we implement an optimal control strategy to assess the efficacy of an anti-tau therapeutic intervention aimed at reducing p-tau levels and mitigating cognitive decline. Our data-driven solutions indicate that while optimal drug administration effectively decreases p-tau concentration, its impact on cognitive decline, as reflected in the ADAS score, remains limited. These findings suggest that targeting p-tau alone may not be sufficient for significant cognitive improvement, highlighting the need for multi-target therapeutic strategies. The integration of mechanistic modelling, advanced parameter estimation, and control-based therapeutic optimization provides a comprehensive framework for improving treatment strategies for AD.
\end{abstract}


\end{frontmatter}

\section{Introduction}

Neurodegenerative diseases are a group of disorders that progress over time and damage the normal functions of the nervous system, such as Alzheimer's disease (AD), Parkinson's disease, and Huntington's disease. Some of the symptoms of these diseases are associated with memory loss, cognitive decline, movement disorders, and impaired motor functions \cite{gao2008}. While the exact causes of many neurodegenerative diseases remain elusive, they are generally thought to result from a combination of genetic, environmental, and lifestyle factors. Research is ongoing to better understand the mechanisms underlying neurodegeneration and to develop effective treatments to prevent it. Each of these neurodegenerative disease conditions has distinct clinical features, but they share some common pathological mechanisms, such as the accumulation of abnormal proteins, mitochondrial dysfunction, and neuroinflammation \cite{peggion2024, rehman2023}. These accumulations of misfolded proteins cause neuronal cell death and the disruption of neural circuits. 

There are two key biomarkers in AD pathology: amyloid-beta (A$\beta$) plaques and tau protein tangles, whose accumulation and interactions drive cognitive decline \cite{bloom2014, gulisano2018}. Mathematical modelling provides a powerful framework for analyzing the progression of neurodegenerative diseases by capturing the dynamics of these pathological biomarkers. In this study, we have developed an ordinary differential equation (ODE)-based model to describe the accumulation and propagation of A$\beta$ and phosphorylated tau (p-tau) in the brain along with Alzheimer’s Disease Assessment Scale (ADAS)-Cognitive Subscale score, a clinical measure of cognitive impairment. Logistic growth rates have been considered for each of the substances \cite{whittington2018, pal2023}, ensuring that their accumulation follows a saturating dynamic constrained by biological limitations. Furthermore, the Michaelis-Menten type kinetics is employed in the p-tau equation to capture the nonlinear effects of amyloid-beta on tau phosphorylation and aggregation, providing insight into its role in the cascade of tau proteins \cite{michaelis1913}. This enzymatic reaction framework allows for a more realistic representation of the biochemical interactions governing disease progression. In addition, it is assumed that the cognitive decline score increases proportionally to the levels of both A$\beta$ and p-tau, reflecting their combined neurotoxic impact on cognitive function.

We have utilised patient data from the Alzheimer’s Disease Neuroimaging Initiative (ADNI) to calibrate our mathematical model \cite{petersen2010}. The ADNI dataset provides longitudinal biomarker measurements, including imaging, cerebrospinal fluid (CSF) markers, and cognitive scores, which serve as valuable inputs for model validation and parameter estimation. The accuracy of these models heavily depends on the estimation of model parameters, which can vary significantly among patients \cite{hao2022}. Traditional parameter estimation techniques often struggle with handling uncertainty and the heterogeneity present in real-world patient data \cite{briggs2012}. Nowadays, machine learning and artificial intelligence automate the discovery of physics principles and governing equations from data \cite{de2020}. 

This work employs two advanced methodologies for parameter estimation: the Bayesian approach and the Physics-Informed Neural Network (PINN) framework \cite{menix2018, raissi2019}. The Bayesian approach enables probabilistic parameter estimation, incorporating prior knowledge and quantifying uncertainty \cite{van2014}, making it well-suited for modelling complex biological systems with noisy and limited data. On the other hand, PINNs leverage deep learning to integrate data-driven insights with the underlying physical laws described by ODEs, offering a flexible and scalable solution for parameter inference. In both approaches, the initial estimations of the parameter values play a crucial role due to the non-convex nature of the loss function. So, we have applied a sequential equation-by-equation algorithm to achieve the initial approximation \cite{hao2022}. We have shown that a wide range of parameter values could be possible for the ODE model while achieving almost the same mean square error for both approaches. 

Beyond parameter estimation, this study explores an optimal control strategy to regulate p-tau accumulation, a key factor in AD progression. The limited success of amyloid-$\beta$-targeting therapies for AD has led to a shift in focus towards the tau protein \cite{congdon2023}. Anti-tau therapies aimed at reducing tau phosphorylation, aggregation, or spreading have emerged as promising treatment strategies \cite{congdon2018, sigurdsson2018, jadhav2019, mummery2023, imbimbo2023}. Despite the fact that targeting tau degradation may have harmful effects, including the failure of the first anti-tau antibody \cite{mullard2021, jabbari2021}, recent studies suggest that reducing tau levels might be beneficial by identifying novel regulators and therapeutic pathways for tau degradation \cite{uliassi2025, hoehne2025}. In addition, optimizing these interventions to maximize therapeutic efficacy while minimizing adverse effects remains a critical challenge. By formulating an optimal control problem, we have investigated potential interventions that minimize tau intervention and cognitive impairment with data-driven pathology. There are two main approaches to solving optimal control problems: Pontryagin’s Maximum Principle (PMP) and the Hamilton-Jacobi-Bellman equation \cite{pontryagin2018, bardi1997}, and we have applied PMP to solve the optimal control problem. The proposed approach provides a data-driven approach based on the tau-targeting antisense oligonucleotide. Mummery et al. observed a declination in CSF p-tau181 in participants treated 24 weeks post-last dose, with a mean percentage change of 56\% from the baseline \cite{mummery2023}. This helps estimate the upper limit of the optimal control set. We have shown that the optimal control model predicts a 55\% reduction in p-tau concentration in the brain, which is a significant match with the controlled treatment patient’s data.

The rest of the article is structured as follows: we first present the preprocessing and integration of ADNI patient data into our framework, followed by the mathematical formulation of the ODE-based neurodegenerative disease model. Next, a detailed discussion of the Bayesian and PINN methodologies for parameter estimation, along with their initial approximation. After that, we consider the optimal control problem to incorporate the anti-tau drug for AD group individuals. Finally, we evaluate the performance of our approach through silico trials and discuss its implications for disease modelling and clinical applications, followed by a summary and future directions.


\section{Data Integration}

AD is currently incurable and causes loss of function and neuronal death over decades. Multiple longitudinal datasets with simultaneous measurements enable innovative medical research by displaying, summarizing, and aggregating data and offering individualized predictions of outcomes. The four consecutive studies (ADNI-1, ADNI-GO, ADNI-2, and ADNI-3) were launched in 2003 by the Alzheimer's Disease Neuroimaging Initiative (ADNI) to analyze AD progression in the brain. The study has collected different biomarker data (adni.loni.usc.edu), such as imaging, cognitive, blood- and CSF-based biomarkers, and genetic data from participants aged between 54 and 90 and monitored for 5-10 years. The dataset contains 1737 individuals' clinical data for applying quantitative observations for AD progression. Based on the standard outline ADNI protocol, these patients are classified as cognitively normal (CN), late mild cognitive impairment (LMCI), mild cognitive impairment (MCI), early mild cognitive impairment (EMCI), and AD. This paper considers the ADNI dataset for the CN, LMCI, and AD subgroups of the participants' CSF beta-amyloid peptide and p-tau at baseline and intervals every six months for up to ten years. We have considered only the data for the individuals with clinical follow-up visits when all three data we are interested in (A$\beta$, p-tau, and ASAD13) are available, as the data set contains some missing values in either of these entries.
 
\begin{figure}[ht!]%
\centering
\includegraphics[width=\textwidth]{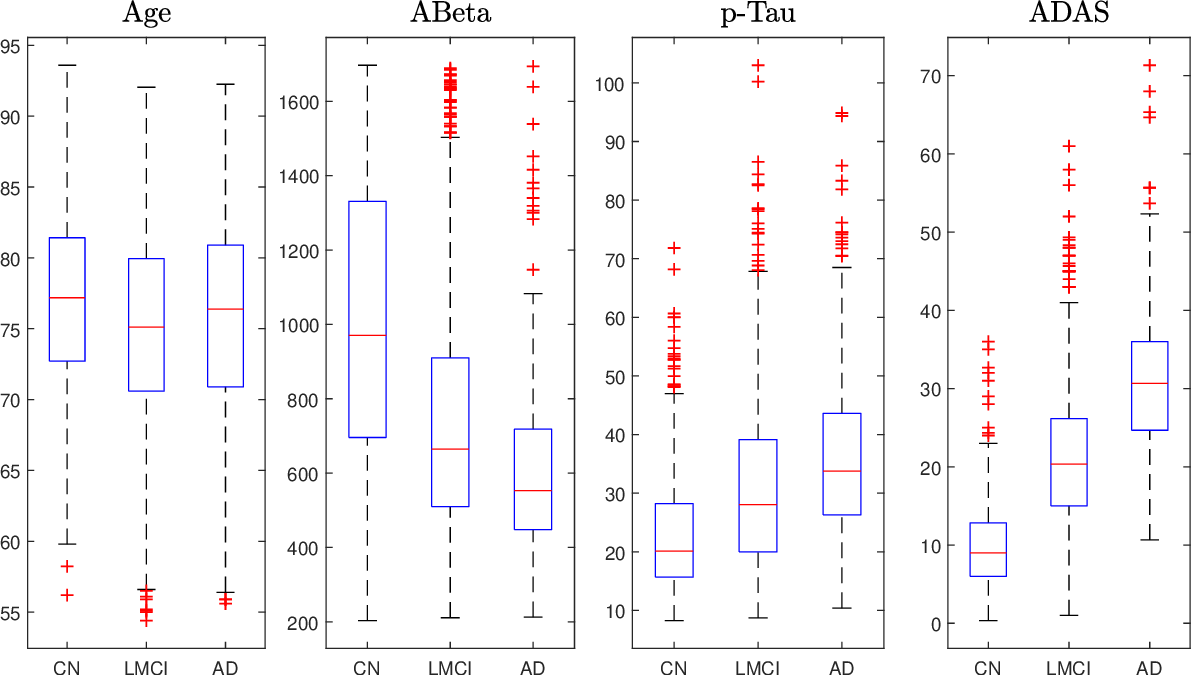}
\caption{ (Color online) Box plots for the different groups of patient data. }\label{FigB1}
\end{figure}

Figure \ref{FigB1} represents the box plot for the patient data for CN, LMCI, and AD subgroups, including their age. The age distributions show that the median age for AD group individuals is higher than the other group individuals. This is because most people seek AD diagnosis later in life, yet some become LMCI or even CN. In addition, the box plot for the age distributions for the CN shows the central tendency; however, the LMCI and AD group individuals are right- and left-skewed, respectively. Therefore, the late mild disease symptoms appear in patients of lower age compared to normal and AD patients. Regarding the biomarkers' concentrations, the amyloid-beta concentrations for the CN group individuals are higher than those of the LMCI and AD groups [see Fig. \ref{FigB1}], which matches with other studies \cite{sjogren2000, andreasen2001, hansson2006}. On the other hand, the concentrations of p-tau and ADAS13 for the CN group individuals are lower than the box plots of the CN group individuals for age, amyloid-beta, p-tau, and ADAS13 show that longer whiskers on the top and medians are always closer to the first quartile. It is true that the decrease of beta-amyloid increases the p-tau levels in CSF \cite{sunderland2003}. Overall, the median and interquartile range for amyloid-beta for the CN group is more than the other groups, and the opposite holds for the p-tau and ADAS13.

\section{Mathematical Model}

In research on neurodegenerative disorders such as AD, mathematical models are essential to comprehending the complex dynamics of the illness. These models aim to illustrate the biological processes, relationships, and variables underlying AD. They also help in understanding the influence of various variables on illness outcomes and forecasting the results of prospective therapies. This paper considers three clinical biomarkers for AD: amyloid-beta, tau protein, and cognitive impairment. In AD, the soluble and plaque-forming type of amyloid beta is the first component in the pathophysiological brain network. It encourages aberrant tau protein phosphorylation, resulting in neurodegeneration and cognitive impairment through widespread disruption of the brain's neural network. It is believed that the imbalance between the production and clearance of A$\beta$ peptides results in the accumulation of A$\beta$ in the brain, forming amyloid plaques. The accumulation of A$\beta$ peptides increases their aggregation into oligomers, fibrils, and extracellular plaques. It is widely accepted that the accumulation of amyloid-beta initiates a cascade of events that lead to the phosphorylation of tau protein, contributing to the neurodegenerative process seen in AD \cite{zheng2002}. The interplay between amyloid-beta and tau is crucial to understanding the pathology and progression of the disease. Furthermore, cognitive decline in AD is strongly associated with the accumulation of tau protein, leading to neurofibrillary tangles (NFTs) \cite{ohm2021}. Hyperphosphorylated tau aggregates into paired helical filaments and eventually forms NFTs that disrupt the normal architecture of neurons.

In modelling the dynamics of A$\beta$ and p-tau in a complex geometry such as the brain, a partial differential equations (PDEs) framework is often used to capture their spatial distribution and temporal evolution, incorporating diffusion and reaction terms. However, machine learning advances promising PDE research by learning coordinate systems and reduced-order models to make PDEs more amenable to analysis \cite{brunton2024}. On the other hand, under the assumption of spatial homogeneity or when considering a well-mixed compartment (e.g., a brain region with negligible concentration gradients), the PDE system can be reduced to a set of ODEs. In this work, such reduction assumes that diffusion is rapid enough to homogenize concentrations compared to the aggregations of A$\beta$ and p-tau in the brain. The resulting ODE model captures their temporal dynamics, allowing for tractable mathematical analysis and parameter estimation while still providing valuable insights into AD progression. Considering these factors, we model the dynamics of amyloid-beta, p-tau, and the cognitive disruptions as:
\begin{equation}{\label{TM}}
    \begin{aligned}
        \frac{du}{dt} &= r_{u}u - b_{u}u^2 \equiv f_{1}(u),\\
        \frac{dv}{dt} &= r_{v}v - b_{v}v^2 + \frac{s_{u}uv}{a_u+h_uu} \equiv f_{2}(u,v),\\
        \frac{dw}{dt} &= r_{w}w - b_{w}w^2 + c_{u}uw + c_{v}vw \equiv f_{3}(u,v,w),
    \end{aligned}
\end{equation}
with positive initial conditions $u(t_{0}) = u_{0}$, $v(t_{0}) = v_{0}$, and $w(t_{0}) = w_{0}$. Here, we have considered the logistic growth in each of the biomarkers, and the rate of cognitive decline is assumed to be proportional to the rate at which amyloid-beta and phosphorylated tau accumulate. In addition, Michaelis-Menten-type biochemical kinetics are employed in the p-tau equation to see the influence of amyloid-beta in the cascade of tau proteins \cite{michaelis1913}. Furthermore, all the parameters involved in this model are positive and will be estimated based on the ADNI dataset, including the initial conditions. By incorporating the patient data into the model, we aim to refine the parameter estimations, ensuring that the predicted trajectories closely align with observed clinical trends. Integrating real-world data helps improve the model’s predictive accuracy, making it a valuable tool for understanding disease progression and treatment response. In the subsequent sections, we describe the transformation process in detail, outlining how the parameters of the system of ODEs are estimated using statistical and machine learning techniques.
 
\section{Parameter Estimation}

Here, we integrate the patients' data into the model and estimate the parameters. Before proceeding further, we write the system of ordinary differential equations (\ref{TM}) in the matrix form:
\begin{equation}{\label{SOE}}
    \frac{d\mathbf{x}}{dt} = \mathbf{F}(\mathbf{x},\alpha),
\end{equation}
where $\mathbf{x} = (u,v,w)^{T}$ and $\mathbf{F} = [f_{1}(\cdot), f_{2}(\cdot), f_{3}(\cdot)]^{T}$. Here, $\alpha$ is the set of parameters to be estimated. In this case, we have considered the initial conditions as the parameters. These parameters can be estimated using the imported data $\widetilde{\mathbf{x}}$ for A$\beta$, p-tau, and ADAS by minimizing the sum of the least square error between the computational data and the observed data at the time knots:
\begin{equation}{\label{OPSOE}}
    \min_{\alpha}\sum_{i}||\mathbf{x}(t_i;\alpha) - \widetilde{\mathbf{x}}(t_i)||_{2}^{2}.
\end{equation}
This optimization problem is a non-convex optimization on a higher-dimensional parameter space, and hence, the problem is sensitive to the initial approximation to find a good local minimum. Because the ODE model is a natural cascade model, we split the optimization problem into three problems and approximated the parameters equation-by-equation. In particular, we first estimate the parameters in the A$\beta$ equation by using the imported A$\beta$ data ($\widetilde{u}$) to solve the optimization problem:
$$\min_{r_u, b_u, u_0}\sum_{i}[u(t_i;r_u,b_u) - \widetilde{u}(t_i)]^{2}~~\mbox{with}~u(t_0) = u_0.$$
After estimating the parameters of the A$\beta$ equation, we repeat the process for p-tau and then the cognitive decline equation. This allows us to have a better initial approximation than random initialization. The algorithm of the equation-by-equation approach is as follows (\textbf{Algorithm-I}):
\begin{itemize}
    \item Step 1: Input the patients' biomarker data points as $\widetilde{\mathbf{x}}(t_i)$.
    \item Step 2: Solve the optimization problem 
    $$\min_{r_u, b_u, u_0}\sum_{i}[u(t_i;r_u,b_u) - \widetilde{u}(t_i)]^{2}~~\mbox{with}~u(t_0) = u_0$$
    to obtain a local minimizer $r_u^0, b_u^0$, and $u_0^0$.
    \item Step 3: Fix $r_u^0, b_u^0$, and $u_0^0$ and solve 
    $$\min_{r_v, b_v, s_u, a_u, h_u, v_0}\sum_{i}[v(t_i;r_v, b_v, s_u, a_u, h_u) - \widetilde{v}(t_i)]^{2}~~\mbox{with}~v(t_0) = v_0$$ 
    to obtain the parameters $r_v^0, b_v^0, s_u^0, a_u^0, h_u^0$, and $v_0^0$.
    \item Step 4: Solve the following optimization problem:
    $$\min_{r_w, b_w, c_u, c_v, w_0}\sum_{i}[w(t_i;r_w, b_w, c_u, c_v) - \widetilde{w}(t_i)]^{2}~~\mbox{with}~w(t_0) = w_0$$ 
    to obtain the parameters $r_w^0, b_w^0, c_u^0, c_v^0$, and $w_0^0$.
\end{itemize}

After determining the initial approximation, we apply the Bayesian technique to obtain statistical information about the parameters based on the combined system's optimization problem (\ref{OPSOE}). In addition, we apply a machine learning technique, namely a physics-informed neural network, to predict the parameter set with the least amount of error in data and given physics.

\subsection{Bayesian approach}

The Bayesian approach offers a novel and powerful framework for estimating the parameter values of an ODE model describing Alzheimer's disease progression. The traditional frequentist methods provide point estimates and rely on asymptotic assumptions. In contrast, the Bayesian framework captures parameter uncertainty by generating posterior distributions, enabling a more robust quantification of variability and confidence in parameter estimates. This approach is based on the prior and likelihood distributions of parameters, which we describe here. In this method, the parameter set $\alpha$ is treated as a random variable, and the goal is to find the posterior distribution $\theta(\alpha|\mathbf{x})$ of the parameters. For a given dataset $\mathbf{x}$, the posterior distribution gives the probability density for the values of $\alpha$. In this case, the Bayes' formula gives the posterior density as
$$\theta(\alpha|\mathbf{x}) = \frac{l(\mathbf{x}|\alpha)p(\alpha)}{\int l(\mathbf{x}|\alpha)p(\alpha)d\alpha},$$
where $l(\mathbf{x}|\alpha)$ is the likelihood that contains the measurement error and $p(\alpha)$ is the prior distribution of the parameters \cite{solonen2006}. The function $l(\mathbf{x}|\alpha)$ gives the probability density of the data $\mathbf{x}$ for the given parameter $\alpha$. 

The integral in the denominator is a normalization constant used to normalize the posterior distribution. This approach estimates the parameters in a fully probabilistic sense, and the challenging part of maximizing the posterior distribution is working with the normalization integral, as it can not be calculated analytically. However, the MCMC methods can give the statistical inference for the parameters $\alpha$ without explicitly calculating the normalization constant. The MCMC method generates a sequence of random samples $\alpha_{1}, \alpha_{2},\ldots,\alpha_{m}$, whose distribution asymptotically approaches the posterior distribution as $m$ increases. Here, the Monte Carlo method generates the random number, and the sequence of samples generated at each new point $\alpha_{n+1}$ depends on the previous sample $\alpha_{n}$.

The Metropolis algorithm is one of the most widely used MCMC algorithms introduced in 1950 in statistical physics \cite{metropolis1953}. The algorithm generates parameter values from a proposal distribution and then accepts or rejects based on the proposed value. The algorithm is as follows (\textbf{Algorithm-II}):
\begin{itemize}
    \item Step 1: Initialize a starting point $\alpha_{1}$. Set $\alpha_{\mbox{old}} = \alpha_{n}$ and $\mbox{Chain}(n) = \alpha_{n}$ with $n=1$.
    \item Step 2: Choose a new parameter set $\alpha^{*}$ from the prior distribution, which may depend on the previous point of the chain, i.e., $\alpha^{*} ~ \mathtt{\sim} ~p(\alpha_{\mbox{old}})$.
    \item Step 3: Accept the new parameter set $\alpha^{*}$ with the probability 
    \begin{equation*}
        \mathcal{P}(\alpha_{n},\alpha^{*}) = \mbox{min}\bigg{(}1, \frac{l(\mathbf{x}|\alpha^{*})p(\alpha^{*})}{l(\mathbf{x}|\alpha_{n})p(\alpha_{n})} \bigg{)}.
   \end{equation*}
   If accepted, set $\mbox{Chain}(n+1) = \alpha_{n+1} = \alpha^{*}$ and $\alpha_{\mbox{old}} = \alpha^{*}$. \\
   If rejected, set $\mbox{Chain}(n+1) = \alpha_{n+1} = \alpha_{\mbox{old}}$ and go back to Step 2.
\end{itemize}

\subsection{Physics-informed neural networks approach}

\begin{figure}[ht!]%
\centering
\includegraphics[width=\textwidth]{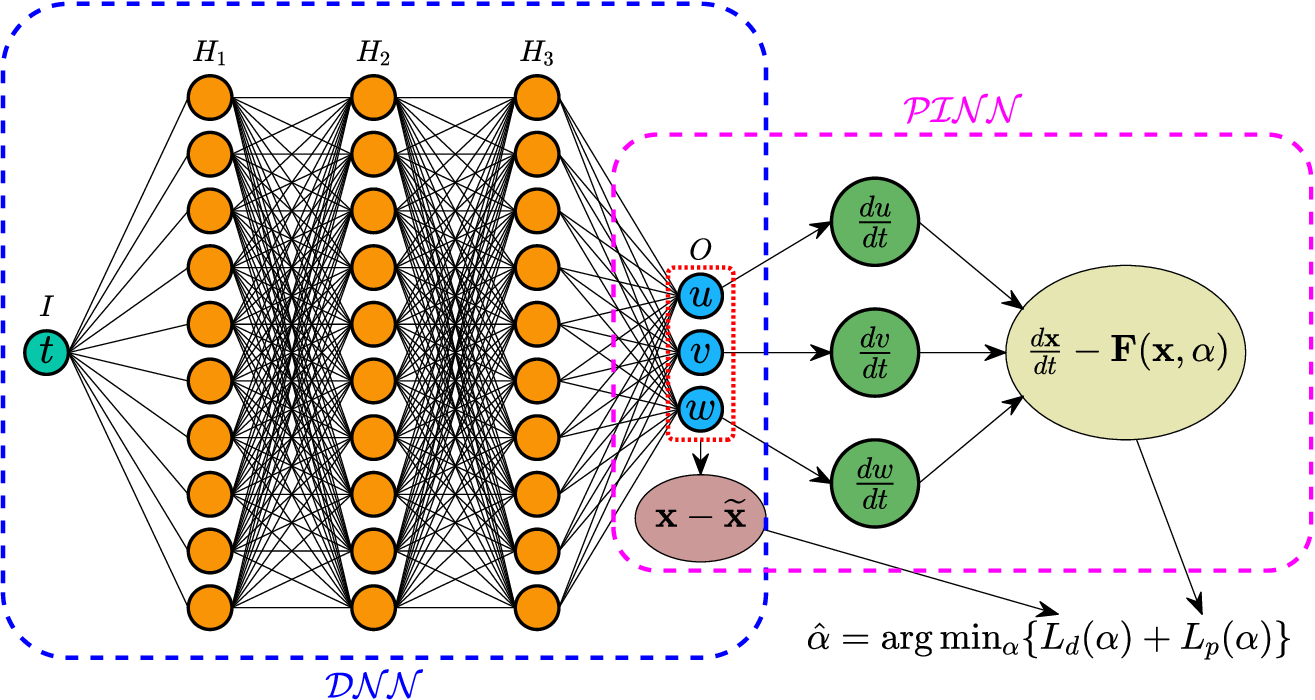}
\caption{ (Color online) The deep neural network employed in the PINNs method consists of input and output layers, denoted as $I$ and $O$, respectively, along with three hidden layers labelled $H_{1}$, $H_{2}$, and $H_{3}$. $L_{d}$ and $L_{p}$ are the mean square losses for the data and the governing physics, respectively.  } \label{FigN1}
\end{figure}

The Bayesian approach helps to estimate the parameter statistics based on the imported ADNI data \cite{shaheen2024}. However, deep learning approaches can estimate those parameter values with the training methodology. For instance, the Physics-informed neural network (PINN) approach combines the data and physics losses in the loss function and optimizes it \cite{raissi2019, cuomo2022}. For this method, we write the problem (\ref{SOE}) as:
\begin{equation*}
    \mathcal{N}(\mathbf{x},\alpha) \equiv \frac{d\mathbf{x}}{dt}-\mathbf{F}(\mathbf{x},\alpha) =  \mathbf{0},~~ t\in[0,T], 
\end{equation*}
with $T>0$. The PINNs approach approximates $\mathbf{x}$ with the help of a deep neural network (DNN), and the parameter set $\alpha$ can be trained by reducing the mean squared error loss:
$$L_{MSE} = \min_{\alpha} \{L_{d}(\alpha)+L_{p}(\alpha)\},$$ where $L_{d}$ is the mean square error loss for the data $\widetilde{\mathbf{x}}$ and $L_{p}$ is the mean square error loss for the governing physics $\mathcal{N}(\mathbf{x},\alpha) = 0$. A schematic diagram of this method is presented in Fig. \ref{FigN1}. The considered ODE system (\ref{TM}) consists of three unknown dependent variables $u$, $v$ and $w$, and an independent variable $t$. Therefore, estimating the parameter values involved in the system (\ref{TM}) requires one neuron in the input layer ($I$) and three neurons in the output layer ($O$) with one or more hidden layers ($H_{i},i=1,2,3,\ldots$) that have the same or different number of neurons each. These hidden layers in a DNN capture progressively more abstract features of the input data. A sufficiently large single hidden layer can approximate any function, but doing so might require an impractically large number of neurons. In addition, training deep networks with many hidden layers used to be difficult due to gradient-related issues, but this has been largely mitigated by advancements like ReLU activation, batch normalization, and residual connections. In this work, as an exemplification of the developed methodology, we have considered three hidden layers with ten neurons in each layer [see Fig. \ref{FigN1}], and the results are given in the numerical simulation section.

\section{Drug control therapies and optimization}

Disease-modifying treatments are crucial for slowing AD progression in the brain \cite{hao2022,pal2023}. Targeting tau protein is more successful in enhancing cognitive function in Alzheimer's disease cases because tau protein's disruption is more strongly associated with dementia than A$\beta$. It is important to note that the development of tau protein treatment is more complex than anti-amyloid therapy. To account for the impact of drugs on tau proteins, we modify the system (\ref{TM}) as:
\begin{equation}{\label{ATE}}
    \begin{aligned}
    \frac{du}{dt} &= r_{u}u - b_{u}u^2,\\
        \frac{dv}{dt} &= r_{v}v - b_{v}v^2 + \frac{s_{u}uv}{a_{u}+h_{u}u} - \phi (t)v,\\
        \frac{dw}{dt} &= r_{w}w - b_{w}w^2 + c_{u}uw + c_{v}vw,
\end{aligned}
\end{equation}
with the initial conditions $u(t_{0}) = u_{0}$, $v(t_{0}) = v_{0}$, and $w(t_{0}) = w_{0}$. Here, the function $\phi (t)$ is the tau protein clearing drug-control function, which is often non-constant over time. This optimal intervention is chosen to minimize both cognitive decline and side effects over the treatment interval $[t_{1}, t_{2}]$, as represented in the following objective function:
\begin{equation}{\label{OFCS}}
    \min_{\phi} J(\phi) := \alpha_{1}v(t_{2}) + \alpha_{2}w(t_{2}) + \int_{t_{1}}^{t_{2}}[\varepsilon (v(t),t)\phi^{2}(t)+w(t)]dt
\end{equation}
with the control set
$$\mathcal{S} = \{\phi \in L^{\infty}([t_{1}, t_{2}]):0\leq \phi \leq \phi^{\max}\}.$$
Here, the term $\varepsilon (v(t),t)\phi^{2}(t)$ represents the side-effect of the anti-tau protein treatment relative to its benefit over time, and, more specifically, the function $\varepsilon(v(t),t)$ depends on tau protein along with the time duration of the treatment. Since the side effects decay with the time of treatment, we assume they decay exponentially in time in the form:
$$\varepsilon (v(t),t) = \epsilon_{0}v(t)e^{-\gamma (t-t_{1})},$$
for some positive constant $\epsilon_{0}$. Our target is to find the optimal control $\phi^{*}$ such that it optimizes the functional $J$, i.e., 
$$J(\phi^{*}) = \min_{\phi \in\mathcal{S}}J(\phi).$$ 
Taking these configurations, the anti-tau therapy control system can be written as
\begin{equation}{\label{SSE}}
    \frac{d\mathbf{x}}{dt} = \mathbf{G}(\mathbf{x}(t),\phi (t)),
\end{equation}
where $\mathbf{x} = [u, v, w]^{T}$ and $\mathbf{G}(\mathbf{x},\phi )  = [f_{1}(u), f_{2}(u,v) -\phi v, f_{3}(u,v,w)]^{T}$. Here, each term depends on the time variable $t$, and they are dropped for notational convenience. Now, we introduce the Hamiltonian for this control problem based on optimal control theory:
$$H(\mathbf{x}(t),\phi(t),\Lambda(t)) = \varepsilon (v(t),t)\phi^{2}(t)+w(t) +\Lambda(t)^{T} \mathbf{G}(\mathbf{x}(t),\phi(t)),$$
where $\Lambda = [\lambda_{1}, \lambda_{2}, \lambda_{3}]^{T}$ is the adjoint vector. After using the Pontryagin’s Maximum Principle, we obtain the system of adjoint equations as:
\begin{equation}{\label{ADJE}}
    \begin{aligned}
        \frac{d\lambda_{1}}{dt} &= -\lambda_{1}(r_{u}-2b_{u}u) - \lambda_{2}\frac{a_{u}s_{u}v}{(a_{u}+h_{u}u)^{2}}  - \lambda_{3}c_{u}w,\\
        \frac{d\lambda_{2}}{dt} &= -\epsilon_{0}e^{-\gamma (t-t_{1})}\phi-\lambda_{2}\left(r_{v}-2b_{v}v +\frac{s_{u}u}{a_{u}+h_{u}u}-\phi\right),\\
        \frac{d\lambda_{3}}{dt} &= -1-\lambda_{3}(r_{w}-2b_{w}w+c_{u}u+c_{v}v),
    \end{aligned}
\end{equation}
with terminal conditions read $\lambda_{1}(t_{2}) = 0$, $\lambda_{2}(t_{2}) = \alpha_{1}$, and $\lambda_{3}(t_{2}) = \alpha_{2}$. In addition, the optimal anti-tau therapy on the interior of the control set can be determined by taking the derivative of the Hamiltonian function with respect to the optimal control input as
\begin{equation*}
    \frac{\partial H}{\partial\phi} = 2\varepsilon \phi  - \lambda_{2}v = 0 \Rightarrow \phi^{*} = 
    \begin{cases}
        ~~ 0&\mbox{if}~\frac{\partial H}{\partial\phi}<0 \\
       ~ \frac{\lambda_{2}v}{2\varepsilon }~&\mbox{if}~\frac{\partial H}{\partial\phi}=0 \\
        \phi^{\max}&\mbox{if}~\frac{\partial H}{\partial\phi}>0 \\
    \end{cases}
\end{equation*}
Therefore, we obtain the optimal control characterization,
\begin{equation}{\label{OCC}}
    \phi^{*} (t) = \min\left[ \phi^{\max}, \max \left\{  0, \frac{\lambda_{2}(t)v(t)}{2\epsilon (t)}\right\}\right].
\end{equation}
As we see, the optimal control problem depends on the state equations (\ref{SSE}) and the adjoint equation (\ref{ADJE}) along with the optimal function $\phi^{*} (t)$. Since the state equations have initial conditions coupled with the adjoint equations, which have terminal conditions, we employ an iterative approach known as the forward-backwards sweep algorithm to solve the optimality system [9]. The algorithm is as follows (\textbf{Algorithm-III}):

\begin{itemize}
    \item Step 1: Input the parameter values and initial conditions for the state variables.
    \item Step 2: Initialize the control $\phi (t)$ as a zero function.
    \item Step 3: Solve the state equation (\ref{SSE}) forward in time using the control $\phi(t)$ to compute $\mathbf{x}(t)$.
    \item Step 4: Solve the adjoint equation (\ref{ADJE}) backward in time using the computed states $\mathbf{x}(t)$ and control $\phi (t)$ to obtain $\Lambda(t)$.
    \item Step 5: Update the control $\phi (t)$ by applying the optimal control characterization (\ref{OCC}) and refine the control function using a convex combination of the previous and updated control.
    \item Step 6: Compute the relative error of states, adjoints and the control. Continue to repeat Steps 3 to 5 until the error is small.
\end{itemize}

\color{black}

\section{Results and Discussions}

This section explains the complex interactions between A$\beta$, p-tau, and ADAS13 scores in AD. The numerical simulations leverage computational models to analyze our integrated ADNI datasets, uncovering patterns and predicting disease progression. As we have discussed, A$\beta$ and p-tau are key biomarkers associated with the pathogenesis of AD, with A$\beta$ plaques and tau tangles contributing to neurodegeneration. In addition, the ADAS13 score is a cognitive assessment tool that quantifies the severity of cognitive impairment in AD patients. Our target is to simulate the dynamics of these biomarkers, their correlation with ADAS13 scores and how changes in A$\beta$ and tau levels impact cognitive decline. This manuscript uses the ADAS score instead of ADAS13 for notational convenience.

\subsection{Initial approximations of the parameters}

\begin{figure}[ht!]%
\centering
\includegraphics[width=0.9\textwidth]{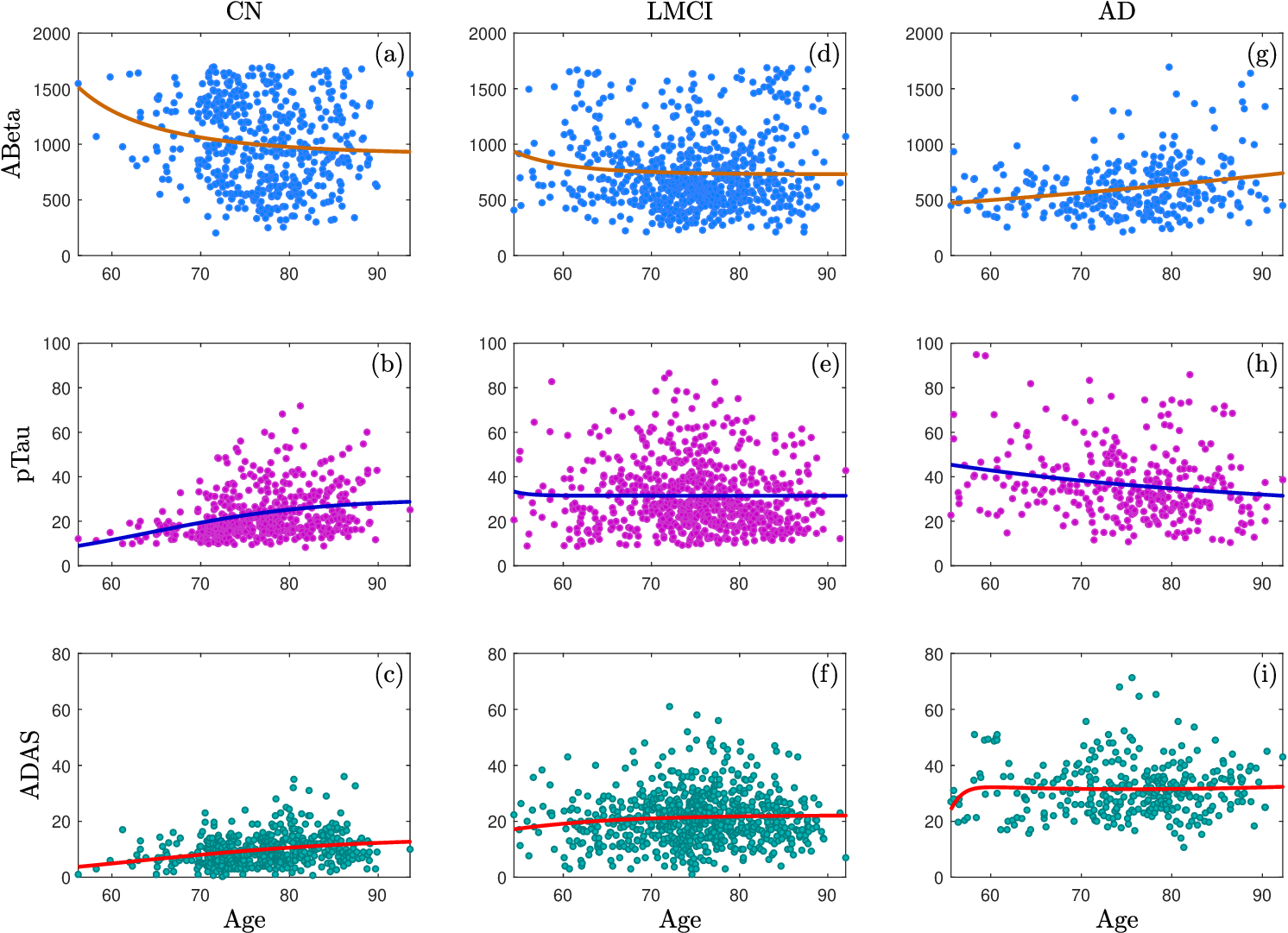}
\caption{ (Color online) The amyloid-beta ($u$), phosphorylated tau ($v$), and cognitive decline ($w$) profiles for the ODE model with the estimated initial approximation for the parameters based on the CN (left-panel), LMCI (middle-panel), and AD (right-panel) patient data using Algorithm-I. }\label{FigI1}
\end{figure}

We first estimate the initial approximation of the parameter values of the ODE model for CN, LMCI, and AD patient subgroups. Algorithm-I estimates the initial approximation of the parameter values, which are given in Table \ref{Tab1}. We use these parameter sets and plot the solutions of the ODE model with the ADNI data in Fig. \ref{FigI1}. We first see the solution behaviour of the ODE model (\ref{TM}) for the CN individuals. The solution corresponding to the amyloid beta shows a decrease in density. This is a part of normal ageing and is linked to the complex interplay between A$\beta$ production, clearance, and deposition in the brain over time. It is often interpreted as A$\beta$ being deposited into plaques in the brain rather than circulating freely in the CSF. In addition, the efficiency of the A$\beta$ clearance mechanisms, such as glymphatic drainage or enzymatic degradation, diminishes with age, potentially leading to increased plaque formation. Therefore, reductions in A$\beta$ in CN ageing are not necessarily indicative of AD but might reflect a natural ageing process that doesn't always result in neurodegeneration. 

\begin{table}[!ht]
\centering
\begin{tabular}{|c|c|c|c|}
\cline{1-4}
 & \multicolumn{3}{|c|}{Value}\\
\cline{2-4}
Parameter & CN group ($n=492$) & LMCI group ($n=704$) & AD group ($n=323$) \\
 \hline
$r_u$ & $0.074$ & $ 0.132 $ & $ 0.0181 $  \\
\hline
$b_u$ & $8.147 \times 10^{-5}$ & $1.811 \times 10^{-4}$ & $ 1\times 10^{-6}$  \\
\hline
$r_v$ & $0.105$ & $0.561$ & $0.0022$  \\
\hline
$b_v$ & $3.514 \times 10^{-3}$ & $0.0178$ & $ 4.5 \times 10^{-4}$  \\
\hline
$s_u$ & $0.188$ & $0.0018$ & $0.95$  \\
\hline
$a_u$ & $8.235$ & $1.56$ & $ 4.397$  \\
\hline
$h_u$ & $23.036$ & $9.554$ & $ 47.65$  \\
\hline
$r_w$ & $4.018 \times 10^{-5}$ & $0.0932$ & $ 1.224 \times 10^{-5}$  \\
\hline
$b_w$ & $4.39 \times 10^{-3}$ & $4.61 \times 10^{-3}$ & $ 0.03$  \\
\hline
$c_u$ & $6.56 \times 10^{-5}$ & $1.02 \times 10^{-5} $ & $ 7.047 \times 10^{-4}$  \\
\hline
$c_v$ & $9.813 \times 10^{-5}$ & $4.43\times 10^{-5}$ & $0.0143$  \\
\hline
$u_0$ & $1514.308$ & $934.177$ & $ 472.06$  \\
\hline
$v_0$ & $8.914$ & $33.347$ & $45.39$  \\
\hline
$w_0$ & $3.723$ & $17.144$ & $24.22$  \\
\hline
\end{tabular}
\caption{Estimated initial approximation for the parameters for the proposed model (\ref{TM}). Here, $n$ is the number of subjects. } \label{Tab1}
\end{table}

On the other hand, the solution corresponding to the p-tau shows a modest increase in the density with age for CN individuals [see the left-middle panel of Fig. \ref{FigI1}]. This reflects age-related changes in neuronal metabolism and tau processing, especially in regions where the first show tau pathology, such as the medial temporal lobe, including the entorhinal cortex. Some primary factors of this gain are low-grade neuroinflammation and oxidative stress. The solution for cognitive decline is also increasing with age for CN individuals [see the left-bottom panel of Fig. \ref{FigI1}] due to a gradual slowdown in processing speed or finding it harder to recall specific details, like names or facts. This is a normal part of ageing and does not usually interfere with daily functioning.

Next, we move to the case of LMCI individuals. Here, the ODE model exhibits the same A$\beta$-related behaviours and cognitive impairment as CN individuals. However, the solution for p-tau becomes almost unchanged with age, but this is not true for CN individuals. The estimated parameter values for these two groups indicate remarkable differences in initial conditions. The initial density for the A$\beta$ for CN individuals is higher than for the LMCI individuals, whereas the opposite happens for the p-tau and ADAS score cases. 

Lastly, we discuss the results for the AD individuals. The estimation of the initial value for ODE corresponding to the A$\beta$ is very low compared to the other two cases. In addition, the concentration of A$\beta$ increases with age, which has not been observed for the CN and LMCI individuals; particularly, the opposite occurs for them. The solution of p-tau shows a decline with age, and the cognitive decline remains constant, except for some ages near the initial condition. Overall, the ODE model (\ref{TM}) predicts the opposite dynamics for AD individuals compared to CN and LMCI individuals by choosing the initial approximation of the parameter values obtained from Algorithm-I.

\subsection{Bayesian statistics of the parameters}

This section presents a detailed statistical analysis of the parameter values derived from the estimated initial approximations obtained in the previous section, providing a comprehensive overview of their distribution and variability. For the Bayesian approach, we employ truncated normal distributions as priors for the parameter values, ensuring that the mean of each distribution corresponds to the initial estimation of the respective parameter. Since the parameters are non-negative, the lower bounds of the distributions are set to zero, effectively constraining the priors to the non-negative domain. To initialize the parameter estimation process, we perform ten thousand simulations to generate the initial Markov chain. Subsequently, leveraging the Bayesian methodology, we conduct fifty thousand additional simulations to refine the posterior distributions and collect robust statistical information for the parameters. This iterative process ensures that the resulting estimates are both accurate and reflective of the underlying probabilistic model. Table \ref{Tab2} summarizes the estimated parameter values for the last five thousand simulations for all three groups. In each group, the number $n$ represents the number of subjects, and the parameter values are given in mean and standard deviation form. 
\begin{table}[!ht]
\centering
\begin{tabular}{|c|c|c|c|}
\cline{1-4}
 & \multicolumn{3}{|c|}{Estimations}\\
\cline{2-4}
Parameters & CN group ($n=492$) & LMCI group ($n=704$) & AD group ($n=323$) \\
 \hline
$r_u$ & $0.0699\pm 0.0214 $ & $0.2148 \pm 0.0972  $ & $0.0657\pm 0.0531 $  \\
\hline
$b_u$ & $(7.761 \pm 2.034) \times 10^{-5}$ & $(2.897 \pm 1.28) \times 10^{-4}$ & $(9.369 \pm 9.161) \times 10^{-5}$  \\
\hline
$r_v$ & $0.1295\pm 0.0545$ & $0.5966\pm 0.101$ & $0.4313\pm 0.2823$  \\
\hline
$b_v$ & $0.005\pm 0.0023$ & $0.0193\pm 0.0033 $ & $0.0126 \pm 0.0077$  \\
\hline
$s_u$ & $0.1925 \pm 0.091$ & $0.0769\pm 0.0578$ & $0.9675 \pm 0.1017$  \\
\hline
$a_u$ & $9.2253\pm 5.3997$ & $6.2679\pm 4.6384$ & $7.2027\pm 4.7926$  \\
\hline
$h_u$ & $25.2216\pm 8.826$ & $14.014\pm 7.4597$ & $46.3177\pm 8.9833$  \\
\hline
$r_w$ & $0.0432\pm 0.336$ & $0.0881\pm 0.0608$ & $0.0722 \pm 0.053$  \\
\hline
$b_w$ & $0.0205\pm 0.0075$ & $0.0206\pm 0.0069$ & $0.0337\pm 0.0085 $  \\
\hline
$c_u$ & ($5.763\pm 4.815)\times 10^{-5}$ & $(2.518\pm 1.901)\times 10^{-4}$ & $(7.754\pm 4.877)\times 10^{-4}$  \\
\hline
$c_v$ & $0.0052\pm 0.0033$ & $0.0053\pm 0.0039$ & $0.0146\pm 0.0075$  \\
\hline
$u_0$ & $1500.2\pm 118.48$ & $949.47\pm 191.92$ & $459.92\pm 56.58$  \\
\hline
$v_0$ & $7.965\pm 3.183$ & $36.71\pm 11.79$ & $49.63\pm 9.18 $  \\
\hline
$w_0$ & $9.206\pm 6.85$ & $21.42\pm 10.97$ & $27.34\pm 10.93$  \\
\hline
\end{tabular}
\caption{Estimated parameter values (mean $\pm$ standard deviation) for the proposed model (\ref{TM}) using the last five thousand posterior Bayesian samples out of fifty thousand samples.} \label{Tab2}
\end{table}

\begin{figure}[ht!]%
\centering
\includegraphics[width=\textwidth]{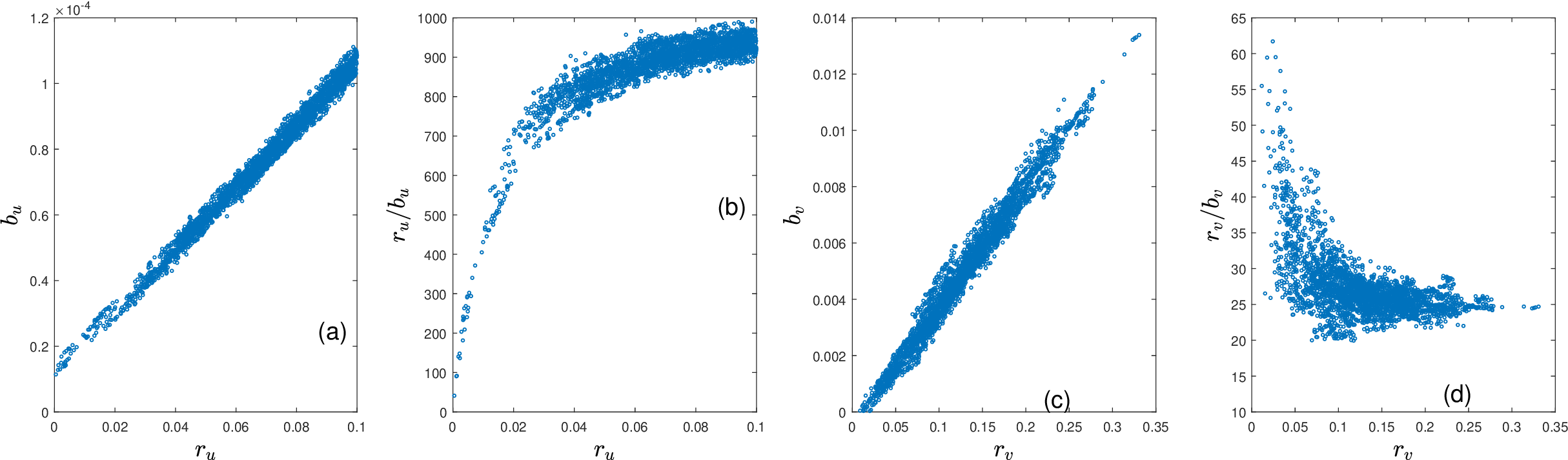}
\caption{ (Color online) Relation between the estimated parameters using Bayesian posterior samples.}\label{FigD1}
\end{figure}

Comparing the initial and estimated parameter values, we observe that a significant
change occurs, indicating that the underlying biological processes governing A$\beta$ accumulation influence the tau accumulation and cognitive declines. Additionally, some of the parameters are correlated with others, suggesting interdependent relationships that shape the progression of A$\beta$ deposition. We first present the correlations between the parameters $r_{u}$ and $b_{u}$ for the CN group individuals in Fig. \ref{FigD1}(a).
In the model formulation, we have considered the temporal evolution of the A$\beta$ as the logistic growth of the form $r_{u}u -b_{u}u^{2}$, which can be converted into the form $r_{u}u[1 -u/(r_{u}/b_{u})]$. In this form, the parameter $r_{u}$ represents the intrinsic growth rate and the factor $r_{u}/b_{u}$ represents the carrying capacity of A$\beta$. Here, the parameter $r_{u}$ is linearly correlated with the parameter $b_{u}$; however, the intrinsic growth rate and the carrying capacity follow a logarithmic relation. This correlation arises because biological constraints, such as enzyme saturation, microglial clearance efficiency, and available aggregation sites, impose limits on A$\beta$ accumulation. A logarithmic relationship implies that beyond a certain threshold of the growth rate, further increases in growth rate lead to diminishing returns in carrying capacity, suggesting a saturation effect in A$\beta$ deposition.

In contrast, the relationship between the growth rate and carrying capacity of p-tau aggregates often follows an exponential decay correlation [see Fig. \ref{FigD1}(d)], meaning that as carrying capacity increases, the growth rate declines rapidly. This exponential decay correlation suggests that interventions targeting tau propagation should focus on early-stage inhibition when the growth rate is still high, as later-stage accumulation becomes less responsive to therapeutic modulation. It further reflects a fundamental difference in tau pathology compared to A$\beta$ aggregation. Unlike A$\beta$, which follows a nucleation-dependent polymerization mechanism, tau pathology spreads primarily through templated misfolding and prion-like propagation, where existing tau fibrils induce the misfolding of soluble tau proteins in neighbouring cells \cite{mudher2017}. Therefore, understanding these distinct relationships between growth dynamics and carrying capacity in A$\beta$ and tau aggregation is crucial for developing stage-specific treatments for neurodegenerative diseases. Furthermore, the rest of the parameters do not show a correlation with others. Except for some quantitative differences, the same scenarios occur for the LMCI and AD group individuals.

\begin{figure}[ht!]%
\centering
\includegraphics[width=\textwidth]{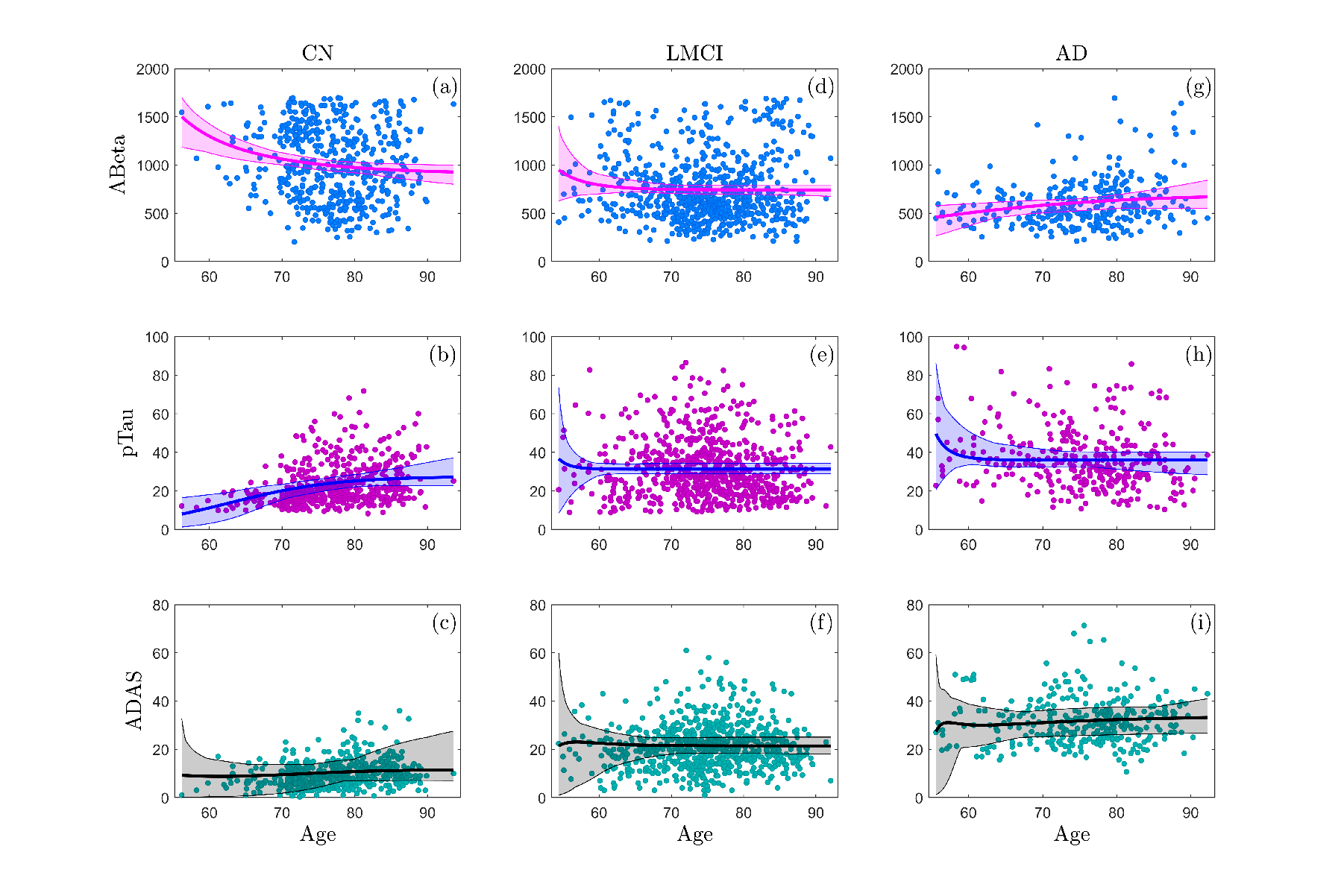}
\caption{ (Color online) The amyloid-beta ($u$), phosphorylated tau ($v$), and cognitive decline ($w$) profiles for the ODE model with the estimated parameter values based on the Bayesian approach. The left, middle, and right panels represent CN, LMCI, and AD patient data, respectively. The shaded regions represent the last five thousand posterior Bayesian samples, whereas the bold curves represent the ODE model solutions evaluated by taking the mean of the parameter values.}\label{FigC1}
\end{figure}

Figure \ref{FigC1} depicts the evolution of the A$\beta$, p-tau, and ADAS densities in time. The solid curve in each plot represents the solution of the ODE model when the parameter values are set to the mean values listed in Table \ref{Tab2}, providing a baseline trajectory of disease progression. The shaded regions illustrate the variability in model solutions when considering the last five thousand posterior Bayesian samples, capturing the range of possible outcomes given parameter uncertainty. This is obtained by finding the maximum and minimum values of the solutions each time when the parameter values are considered from posterior Bayesian samples. This result indicates that the model's solution profile remains consistent with the initial parameter estimates. However, it also highlights that all three biomarkers- A$\beta$, p-tau, and ADAS- can exhibit diverse trajectories depending on initial conditions. This variability is largely attributed to the scarcity of patient data in the early age range, which limits the precision of model predictions in the initial stages of disease progression. Consequently, further data collection in younger at-risk populations could refine these estimates and improve predictive accuracy for early-stage Alzheimer's disease.

\subsection{Parameter estimation using PINNs approach}

\begin{figure}[ht!]%
\centering
\includegraphics[width=0.6\textwidth]{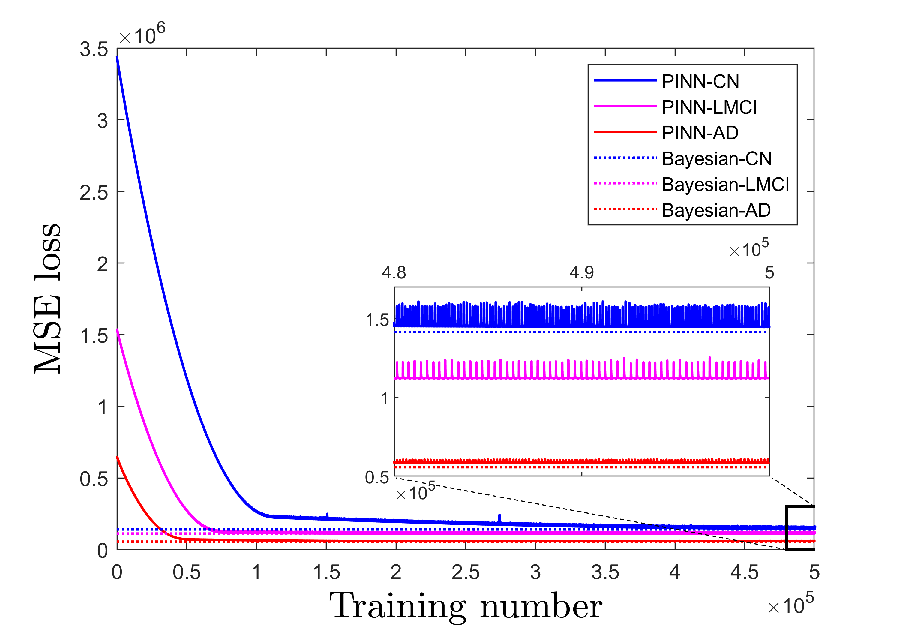}
\caption{ (Color online) Mean-square error (MSE) loss in the PINNs training process for the ODE model for CN, LMCI, and AD group individuals. The dotted lines are MSE loss for the ODE model corresponding to the mean values of the estimated parameters given in Table \ref{Tab2}.}\label{FigF1}
\end{figure}

Here, we estimate the parameter values using the PINNs method. The initial parameter value estimations have been considered the initial approximation for the PINNs method. We have considered one input layer, three hidden layers with ten nodes each, and three output layers, with the weights and biases initialized to zero. In addition, the learning rate and error tolerance are considered as $10^{-3}$ and $10^{-8}$, respectively. The neural network is trained using the mean square loss function, which quantifies the discrepancy between the predicted values from the ODE solution and the observed ADNI patient data. During each training iteration, the mean squared error (MSE) is calculated by solving the ODE model using the estimated parameters and comparing the results with the patient data. We plot the mean square loss in Fig. \ref{FigF1} during the training process.
The figure illustrates that the error decreases as training progresses, demonstrating the network's ability to learn the underlying parameter values. However, towards the later stages of training, the loss function exhibits tiny oscillations, which indicate convergence challenges and are sensitive to the learning rate. Despite these fluctuations, the training process ultimately converges, leading to the final parameter estimations. Figure \ref{FigF1} shows a significant decrease in error loss during the initial stages of training (e.g., when the number of training iterations is less than $10^5$ for CN individuals). However, as training progresses, the error loss decreases at a much slower rate compared to the initial phase. Notably, two small spikes in the MSE loss for CN individuals appear around $1.5 \times 10^5$ and $2.65 \times 10^5$ iterations. This behaviour aligns with the double descent phenomenon, where the deep neural network overfits before improving its generalization performance \cite{kempkes2025}. Eventually, the MSE loss continues to decrease when training continues. Standard Bayesian approaches in cognitive science and brain studies are promising, but they face empirical challenges in decision-making problems, among others \cite{favela2023, Griffiths2024}. Indeed, experimental evidence suggests deviations from rational, probabilistic thinking and highlights the limitations of a purely rational Bayesian framework.

Furthermore, we plot the MSE obtained for the Bayesian approach in Fig. \ref{FigF1}. It shows that the MSE loss for each group for the PINNs method converges to the corresponding MSE loss of the Bayesian method, indicating that both approaches achieve comparable levels of accuracy in minimizing the error. This suggests that PINNs can effectively approximate the underlying dynamics of the system while maintaining consistency with the Bayesian inference results. Nevertheless, Table \ref{Tab3} summarizes the estimated parameter values for all the substances with different groups of individuals, highlighting the variations in parameter estimation across groups. Comparing Tables \ref{Tab2} and \ref{Tab3} gives a wide range of parameter values that could be possible for the ODE model while achieving almost the same MSE. This observation underscores the potential non-identifiability of certain parameters, where multiple parameter sets yield similar model performance in terms of error minimization. These findings emphasize the importance of incorporating additional constraints or prior knowledge to refine parameter estimation and improve interpretability in complex biological systems.

\begin{table}[!ht]
\centering
\begin{tabular}{|c|c|c|c|}
\cline{1-4}
 & \multicolumn{3}{|c|}{Value}\\
\cline{2-4}
Parameter & CN group ($n=492$) & LMCI group ($n=704$) & AD group ($n=323$) \\
 \hline
$r_u$ & $0.142$ & $ 0.14$ & $ 0.0691$  \\
\hline
$b_u$ & $1.42 \times 10^{-4}$ & $1.873\times 10^{-4} $ & $ 1.129\times 10^{-4}$  \\
\hline
$r_v$ & $0.078$ & $0.472 $ & $ 0.0093$  \\
\hline
$b_v$ & $0.0036$ & $ 0.153$ & $ 7.2\times 10^{-4}$  \\
\hline
$s_u$ & $0.161$ & $ 0.0918$ & $ 0.842$  \\
\hline
$a_u$ & $8.424$ & $ 1.472$ & $ 4.434$  \\
\hline
$h_u$ & $23.223$ & $ 9.405$ & $47.76$  \\
\hline
$r_w$ & $1.278\times 10^{-4}$ & $ 0.0562$ & $ 0.0079$  \\
\hline
$b_w$ & $0.0132$ & $ 0.026$ & $ 0.0324$  \\
\hline
$c_u$ & $1.277\times 10^{-4}$ & $6.539\times 10^{-4} $ & $ 0.0017$  \\
\hline
$c_v$ & $1.272\times 10^{-4}$ & $1.953\times 10^{-4} $ & $ 1.6\times 10^{-4}$  \\
\hline
$u_0$ & $1004.4$ & $ 748.2$ & $605.1$  \\
\hline
$v_0$ & $23.3$ & $ 31.49$ & $ 36.53$  \\
\hline
$w_0$ & $9.86$ & $21.25 $ & $ 31.54$  \\
\hline
\end{tabular}
\caption{Estimated parameter values for the proposed model (\ref{TM}) using PINNs training.} \label{Tab3}
\end{table}

\subsection{Optimal control}

The effects of drugs on a patient can vary significantly depending on the dose administered. High doses are typically used when a strong and rapid therapeutic response, such as acute or severe symptoms. However, they may increase the risk of side effects or toxicity, particularly if the patient has underlying health issues. In contrast, low doses are often prescribed to minimize adverse effects, especially during initial treatment phases or for long-term maintenance. Therefore, the choice of dose must be tailored to the individual’s condition, response to treatment, and overall health, often requiring close monitoring and adjustment by the healthcare provider. For the considered case, the control function $\phi$ signifies the drug dosage of anti-tau protein for AD patients.


Now, we estimate the parameter $\phi^{\max}$ involved in the optimal control. Based on the tau-targeting antisense oligonucleotide ($MAPT_{RX}$), there are four groups: low dosages to high dosages injections \cite{mummery2023}. It has been observed that CSF p-tau181 continued to decline in participants treated with $MAPT_{RX}$ in a 60mg monthly group 24 weeks post-last dose, with the mean percentage change of 56\% from the baseline. As the time scale of our model is the year, we can approximate 24 weeks as 0.462 years. In this case, we consider 
$$\frac{dv}{dt}=-\phi^{\max}v\Rightarrow v(t) = v(0)e^{-\phi^{\max}t}.$$
Accordingly, we have $$\phi^{\max}=-\frac{\ln{(0.44)}}{0.462} = 1.779.$$

\begin{figure}[ht!]%
\centering
\includegraphics[width=0.7\textwidth]{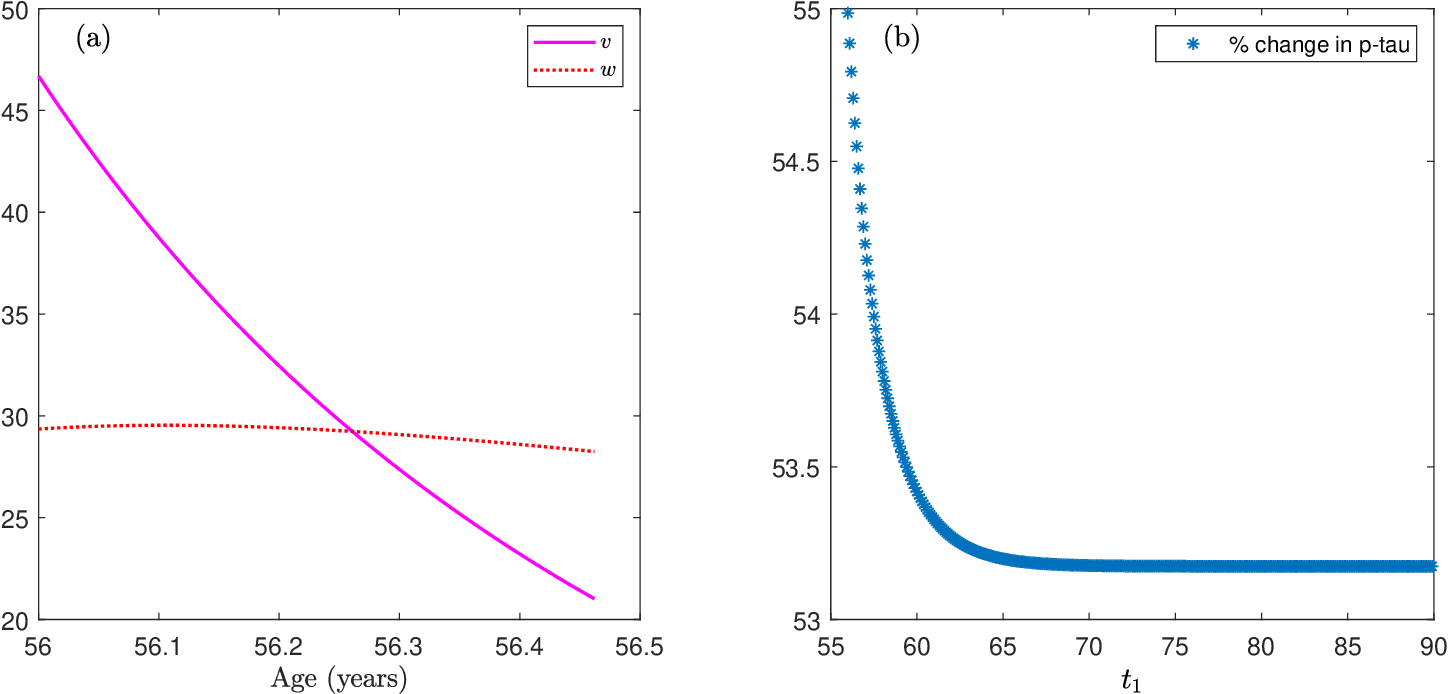}
\caption{ (Color online) (a) Solution of the control problem in the presence of optimum control over 24 weeks of treatment and (b) the percentage change in p-tau concentration before and after the treatment for different starting ages ($t_{1}$) of the patients. }\label{FigO}
\end{figure}

We take $\alpha_1 = \alpha_2 = 1$ in the objective function (\ref{OFCS}) along with $\epsilon_{0}=1$ and $\gamma=1$. Now, we solve the control problem (\ref{ATE}) to minimize the minimizer (\ref{OFCS}) for the AD and LMCI group patients. As the control bound parameter $\phi^{\max}$ is obtained for the AD patient, we first see the model outcome for this group of individuals. We have initially considered the starting age of the anti-tau treatment to be $t_{1} = 56$. Figure \ref{FigO}(a) illustrates that the data-driven solution for p-tau concentration decreases during the treatment, dropping the value from 46.69 to 21.02, which corresponds to an approximate 55\% reduction.
This significant decrease highlights the efficacy of the proposed control strategy in regulating p-tau levels, demonstrating a significant agreement between the mathematical model and empirical patient data.

On the other hand, the ADAS score also decreases during treatment, but its overall reduction is less pronounced compared to p-tau concentration. This suggests that while the anti-tau treatment has a direct impact on the biological markers associated with AD progression, its effect on cognitive decline, as measured by ADAS, may require a longer treatment duration or a combination with other therapeutic interventions. Furthermore, we explore the influence of the initial treatment age on the reduction in p-tau concentration. Figure \ref{FigO}(b) presents the percentage change in p-tau levels before and after the drug is administered for different starting ages. The results indicate a nonlinear, particularly exponential decay, relationship between the age at which treatment begins and the percentage reduction in p-tau concentration. This finding suggests that initiating treatment at an earlier stage of AD can lead to a more substantial reduction in p-tau accumulation, potentially delaying disease progression. 

\begin{figure}[ht!]%
\centering
\includegraphics[width=0.55\textwidth]{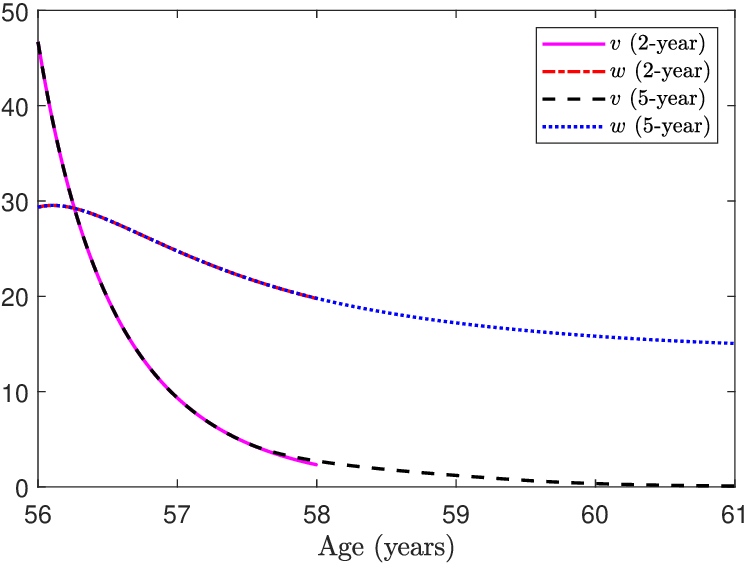}
\caption{ (Colour online) Solutions to the control problem in the presence of optimum control over 2- and 5-year treatment. }\label{FigO5}
\end{figure}

Next, we study the in silico trials when the therapy is administered to AD patients using the optimal control problem. Figure \ref{FigO5} illustrates the reduction of p-tau concentration and the ADAS score for different treatment durations.
The results demonstrate a significant decrease in p-tau concentration within the brain, suggesting that the therapy effectively targets the pathological mechanism underlying AD. However, the ADAS score, which measures cognitive and functional decline, does not exhibit a similarly significant reduction. This discrepancy indicates that while the treatment successfully mitigates the accumulation of toxic tau proteins, it does not necessarily translate into substantial cognitive improvements within the given timeframe. Furthermore, the findings suggest that prolonged medication use does not eliminate the associated side effects, highlighting the need for optimized treatment strategies that balance efficacy with tolerability. This calls for further investigations into alternative dosing regimens, combination therapies, or adjunct interventions that could enhance cognitive outcomes while minimizing adverse effects.

\section{Conclusions and Future Directions}

In this study, we developed and analyzed ODE-based models to describe the progression of AD by incorporating the dynamics of A$\beta$, p-tau, and the ADAS score, a clinical measure of cognitive decline. The models provided insights into the interactions between A$\beta$ and p-tau, demonstrating their roles in disease progression. By leveraging parameter estimation techniques, we calibrated the models using Bayesian inference and PINNs, allowing for data-driven approaches to capturing disease dynamics based on ADNI data. The Bayesian framework enabled the quantification of parameter uncertainties, improving confidence in model predictions, while the PINN approach facilitated learning complex nonlinear relationships directly from data. These complementary methodologies enhanced the predictive power of the models, providing a reliable framework for understanding AD progression. Furthermore, similar to AD, there are other neurodegenerative disorders, such as multiple sclerosis, where anti-tau therapy could be beneficial \cite{hoehne2025}.

Furthermore, we applied an optimal control strategy to investigate the efficacy of an anti-tau drug in reducing p-tau accumulation and its impact on cognitive decline. Our simulations revealed that optimal drug administration significantly lowers p-tau concentration in the brain, which matches clinical observations. However, its effect on the ADAS score remains limited over extended treatment periods, indicating that cognitive improvements do not necessarily follow from biomarker reductions alone. Additionally, silico trials emphasize the importance of early intervention in neurodegenerative disorders and provide valuable insights into the timing of anti-tau therapeutic strategies. Furthermore, prolonged medication use does not eliminate associated side effects, emphasizing the need for tailored treatment strategies. Future research should explore multi-target therapeutic approaches, integrating A$\beta$ and tau-targeting interventions alongside optimized dosing regimens to enhance clinical outcomes. The combination of mechanistic modelling, data-driven parameter estimation, and control-based therapeutic optimization provides a powerful framework for advancing precision medicine in AD. A promising direction for future research involves incorporating astrocyte dynamics into models of amyloid-beta and tau protein interactions \cite{pal2025qb}, which could offer a more comprehensive understanding of AD progression and potential strategies for disease control. 


The Bayesian approach and Artificial Intelligence (AI) intersect significantly, particularly in mathematical modelling, where Bayesian methods provide a powerful framework for dealing with uncertainty and learning from data \cite{van2014, xia2022}. As we have discussed, the Bayesian approach is rooted in Bayes' Theorem, which updates the probability of a hypothesis based on new evidence. It involves the initial beliefs about a model’s parameters before observing data, quantifies the likelihood of the observed data given these parameters, and updates these beliefs based on the observed data. This method provides a natural way to quantify uncertainty in predictions and model parameters, which is valuable in many AI applications where understanding confidence in decisions is crucial - for instance, diagnosing diseases by updating the probability of a disease given new test results and analyzing brain imaging data (e.g., MRI, PET scans) to detect patterns associated with AD. In addition, it allows the incorporation of prior knowledge or expert opinions into the model, which can be particularly useful in scenarios with limited data, e.g., previous knowledge about disease prevalence and test accuracy. The Bayesian approach has a wide range of applications in AI, including classification problems based on the likelihood of various features; autonomous driving systems that demonstrate a strong ability to manage uncertainty, enhance decision-making, and adapt to changing conditions, etc. While AI is increasingly utilized in medical research, including precision medicine \cite{chen2025}, it has specific limitations. One of the biggest challenges is acquiring high-quality, large-scale datasets, which are essential for tackling the issue of poor generalization when applying findings to new patients. Additionally, understanding nonequilibrium phenomena is gaining significance, particularly in the context of AD-modifying therapies. This understanding can help clarify how the brain responds dynamically to treatment interventions over time \cite{pal2025, cruzat2023}. These phenomena highlight that therapeutic effects may not follow a linear or steady progression, with possible transient destabilizations before reaching a new, healthier state. Considering these dynamic shifts, researchers and clinicians can design therapies that harness the brain’s adaptive capacity, potentially improving long-term outcomes despite initial fluctuations.

One of the biggest problems is getting high-quality and large-scale datasets, which can resolve the poor generalization of new patients. In addition, neurodegenerative diseases like Alzheimer’s have complex behaviour and are involved in heterogeneous data, whose longitudinal studies are rare. Furthermore, most AI models need rigorous validation before use in clinical settings, which can be time-consuming and costly at present. However, AI technology advances with large datasets, and hence, its role in neurodegenerative disease modelling can offer new opportunities for improving patient outcomes, such as early detection, personalized treatment, and understanding of disease mechanisms.

\section*{Acknowledgements}
The authors thank the NSERC and the CRC Program for their support. RM also acknowledges the support of the BERC 2022-2025 program and the Spanish Ministry of Science, Innovation and Universities through the Agencia Estatal de Investigacion (AEI) BCAM Severo Ochoa excellence accreditation SEV-2017-0718. This research was partly enabled by support provided by SHARCNET and the Digital Research Alliance of Canada.

\bibliography{References}

\end{document}